# Impact ionization in narrow band gap CdHgTe quantum well with "resonant" band structure

V.Ya.Aleshkin[1,2,a)] A.A.Dubinov[1,3], V.V.Rumyantsev[1,3]

[1]Department of Semiconductor Physics, Institute for Physics of Microstructures RAS, Nizhny Novgorod 603950, Russia
[2]ASGAP, Lobachevsky State University of Nizhny Novgorod, Nizhny Novgorod 603950, Russia
[3]Radiophysical Department, Lobachevsky State University of Nizhny Novgorod, Nizhny Novgorod 603950, Russia

[a)]Author to whom correspondence should be addressed: aleshkin@ipmras.ru

**Abstract**

Impact ionization probabilities were calculated in a CdHgTe quantum well, where the distance between electron subbands is close to the band gap energy. This band structure enables impact ionization with small momentum transfer for electrons in the second subband. The study demonstrates that such processes increase the impact ionization probability by approximately two orders of magnitude compared to the impact ionization probability for electrons in the first subband, for which transitions with small momentum changes are impossible. The probability of single impact ionization during the electron energy loss due to optical phonon emission is estimated. Experimental methods for detecting impact ionization in this structure are discussed.

## 1. Introduction

The phenomenon of impact ionization underlies the operation of many semiconductor devices: avalanche transit-time diodes, avalanche photodetectors, avalanche transistors, etc. [1]. This process is also important for the operation of avalanche QWIPs [2,3] and solar cells [4]. In bulk semiconductors, impact ionization has been studied fairly well (see review [5] and references therein). As a rule, impact ionization is studied in the case of strong electric fields applied to the semiconductor, due to which carriers with high kinetic energy appear, which can participate in this process. However, this phenomenon has been poorly studied in quantum wells, especially in the case when impact ionization occurs not in strong electric fields, but under the influence of optical radiation. It should be noted that in [6] a study was conducted of impact ionization in HgTe quantum wells of critical thickness (band gap close to zero) under the influence of terahertz laser radiation. However, in this work, the probability of impact ionization was not calculated.

In lasers based on narrow-gap HgTe quantum wells [7,8], the energy offsets of the band edges exceed the band gap. In such structures, under optical excitation by photons whose energies are significantly greater than the band gap, impact ionization processes can increase the photocarrier concentration. This can lead to a decrease in the threshold pump energy densities. In the case of current pumping of such structures, impact ionization can also effectively increase the concentration of nonequilibrium electrons and holes in the quantum wells and reduce the threshold current densities.

The aim of this work is to calculate the probability of impact ionization under optical excitation in narrow-band quantum wells. We will consider a quantum well in which the distance between the ground and excited electron subbands is on the order of the band gap. In such structures, impact ionization is possible without a change in electron momentum, when the electron initiating the process is transferred from the excited electron subband to the ground electron subband as a result of impact ionization, and the second electron is transferred from the valence subband to the ground electron subband. The work will show that, due to these processes, the probability of impact ionization involving electrons of the second subband is almost two orders of magnitude higher than the probability of impact ionization involving electrons of the ground electron subband.

It should be noted that impact ionization processes are also possible in bulk semiconductors without changing the electron momentum when the spin splitting of the valence band is close to the band gap [5,9-15]. Sometimes such a band structure is called "resonant" [5,15,16]. In such structures, the rate of impact ionization [9-15] and the rate of Auger recombination involving holes [17-20] increase sharply compared to structures in which such processes are impossible.

## 2. Model

Let's consider the probability of impact ionization, whereby an electron located in one of the quantum subbands of the quantum well's conduction band transfers part of its energy to an electron located in one of the valence subbands. As a result of this process, the electron moves from the valence subband to the lower subband of the conduction band, forming an electron-hole pair. The physical cause of impact ionization is the Coulomb interaction of electrons. Due to the homogeneity of the system in the plane of the quantum well, the potential created by an electron located at a point with radius vector $\mathbf{r}_1$, and by a second electron with radius vector $\mathbf{r}_2$, can be written as:

$$\varphi(\mathbf{r}_1,\mathbf{r}_2)=\varphi(\boldsymbol{\rho}_1-\boldsymbol{\rho}_2,z_1,z_2) \tag{1}$$

where the z-axis is directed along the normal to the quantum well and $\boldsymbol{\rho}_{1,2}$ are the radius vectors characterizing the position of electrons in the quantum well plane. The explicit form of the potentials was given in the works of Rytova [21] and Keldysh [22]. To find the probability of impact ionization, it is necessary to calculate the matrix element of the Coulomb interaction between the electron states before and after scattering. To calculate it, it is necessary to find the wave functions of electrons in the quantum well. To find them, we use the four-band Kane model [23]. In this model the wave functions have the form:

$$\psi_{\mathbf{k},s}(\mathbf{r})=\frac{\exp(i\mathbf{k}\boldsymbol{\rho})}{\sqrt{S}}\mathbf{u}_{\mathbf{k},s}(z) \tag{2}$$

where $\mathbf{k}$ is the electron wave vector, s is the index denoting the subband number and spin state, S is the area of the quantum well, and $\mathbf{u}_{\mathbf{k},s}(z)$ is a column of 8 elements. If the initial states of the electrons are designated by indices 1 and 2, and the final states by indices 3 and 4, then the matrix element of the Coulomb interaction of electrons can be represented as:

$$\begin{aligned}U_{k_1,s_1,k_2,s_2;k_3,s_3,k_4,s_4}&=\frac{e}{S}\Big\{\int dz_1dz_2\varphi_{\mathbf{k}_1-\mathbf{k}_3}(z_1,z_2)\left(\mathbf{u}^+_{k_3,s_3}(z_1)\mathbf{u}_{k_1,s_1}(z_1)\right)\left(\mathbf{u}^+_{k_4,s_4}(z_2)\mathbf{u}_{k_2,s_2}(z_2)\right)-\\&\int dz_1dz_2\varphi_{\mathbf{k}_1-\mathbf{k}_4}(z_1,z_2)\left(\mathbf{u}^+_{k_4,s_4}(z_1)\mathbf{u}_{k_1,s_1}(z_1)\right)\left(\mathbf{u}^+_{k_3,s_3}(z_2)\mathbf{u}_{k_2,s_2}(z_2)\right)\Big\}\delta_{\mathbf{k}_1+\mathbf{k}_2,\mathbf{k}_3+\mathbf{k}_4}=\\&=\frac{V_{k_1,s_1,k_2,s_2;k_3,s_3,k_4,s_4}}{S}\delta_{\mathbf{k}_1+\mathbf{k}_2,\mathbf{k}_3+\mathbf{k}_4}\end{aligned} \tag{3}$$

where $e$ is the electron charge, $\varphi_{\mathbf{q}}(z_1,z_2)=\int d^2\rho\exp(i\mathbf{q}\boldsymbol{\rho})\varphi(\boldsymbol{\rho},z_1,z_2)$ is the Fourier component of the potential, $\delta_{\mathbf{k}_1+\mathbf{k}_2,\mathbf{k}_3+\mathbf{k}_4}$ is the Kronecker symbol, corresponding to the conservation of the total momentum of electrons before and after interaction. In the case electron impact ionization states 1, 3, and 4 are in the conduction band, and states 2 is in the valence band. In the case hole impact ionization states, state 4 is in the conduction band, and states 1–3 are in the valence band. Note that the first term in expressions (3) is the direct Coulomb interaction, and the second term is the exchange interaction. We will assume that the number of electrons in the quantum well is small, so their contribution to the permittivity can be neglected. The explicit form for this case is given in [21,24] (see also the Appendix).

For structures with a "resonant" band structure, when the distance between quantum-confinement subbands is close to the band gap, electron transitions with small changes in the electron wave vector play a major role in impact ionization. For such transitions $\mathbf{q} \to 0$. It is important to note that for $\mathbf{q} \to 0$ $\varphi_{\mathbf{q}}(z_1, z_2) \propto q^{-1}$, what is a consequence of the long-range nature of the Coulomb potential. This circumstance leads to large errors in the numerical calculation of the matrix element of the Coulomb interaction using Eq. (3). To circumvent this difficulty, we note the following. From the explicit form of $\varphi_{\mathbf{q}}(z_1, z_2)$, which is given in the appendix due to its cumbersome nature, we can conclude that for small q, the following expansion is valid:

$$\varphi_{\mathbf{q}}(z_1, z_2) = \frac{1}{q}\left( \frac{2\pi e}{\kappa_1} + q\frac{\partial q\varphi_{\mathbf{q}}(z_1, z_2)}{\partial q} + \frac{q^2}{2}\frac{\partial^2 q\varphi_{\mathbf{q}}(z_1, z_2)}{\partial q^2} + ... \right) \tag{4}$$

where $\kappa_1$ is the permittivity of the quantum well, and derivatives of the function $q\varphi_{\mathbf{q}}(z_1, z_2)$ are taken at q=0. We will now show that the contribution of the first term in (4) to the matrix element tends to zero as $q \to 0$. To do this, consider the first term in curly brackets (3), substituting into it the first term of the expansion (4). Then it takes the form:

$$\frac{2\pi e}{|\mathbf{k}_1 - \mathbf{k}_3|}\int dz_1 \mathbf{u}^{+}_{k_3,s3}(z_1)\mathbf{u}_{k_1,s1}(z_1)\int dz_2 \mathbf{u}^{+}_{k_4,s4}(z_2)\mathbf{u}_{k_2,s2}(z_2) \tag{5}$$

Due to the orthogonality of the functions $\mathbf{u}$ of different subbands for the same $\mathbf{k}$, the integrals over $z_1$ and $z_2$ tend to zero as $\mathbf{k}_1 - \mathbf{k}_3 \to 0$ (note that $\mathbf{k}_1 - \mathbf{k}_3 = \mathbf{k}_4 - \mathbf{k}_2$). Due to the absence of branch points on the real k-axis at the column $\mathbf{u}_{\mathbf{k},s}(z)$, the integrals in (5) can be expanded in a Taylor series in $\mathbf{k}_1 - \mathbf{k}_3$, which have no terms independent of the difference $\mathbf{k}_1 - \mathbf{k}_3$. Therefore, expression (5) tends to zero as $\mathbf{k}_1 - \mathbf{k}_3 \to 0$. Therefore, for sufficiently small q, we can set in formula (3):

$$\varphi_{\mathbf{q}}(z_1, z_2) \approx \frac{\partial q\varphi_{\mathbf{q}}(z_1, z_2)}{\partial q} = g_{\mathbf{q}}(z_1, z_2) \tag{6}$$

Due to its cumbersome nature, the explicit form $g_{\mathbf{q}}(z_1, z_2)$ is given in the appendix to the article. We make one more remark about which q values are considered small. From the expression for $\varphi_{\mathbf{q}}(z_1, z_2)$ given in the appendix and the condition that the electron wave functions rapidly decay outside the quantum well, it is clear that q can be considered small if the condition qd<<1 is satisfied, where d is the quantum well width. In our calculations, expression (6) was used when the condition qd<0.03 was satisfied.

The probability of the impact ionization process caused by an electron with wave vector k can be written as:

$$W(\mathbf{k}_1, s_1) = \frac{2\pi}{\hbar S^2} \sum_{k_2,s_2,k_3,s_3,k_4,s_4} \frac{|V_{k_1,s_1,k_2,s_2;k_3,s_3,k_4,s_4}|^2}{2} \delta_{\mathbf{k_1}+\mathbf{k}_2,\mathbf{k}_3+\mathbf{k}_4} \times$$
$$\delta(\varepsilon_{s_1}(\mathbf{k}_1) + \varepsilon_{s_2}(\mathbf{k}_2) - \varepsilon_{s_3}(\mathbf{k}_3) - \varepsilon_{s_4}(\mathbf{k}_4)) f_{s_2}(\mathbf{k}_2)(1 - f_{s_3}(\mathbf{k}_3))(1 - f_{s_4}(\mathbf{k}_4)) \quad (7)$$

where $\hbar$ is the Planck's constant, $f_{s_j}(\mathbf{k}_j)$ is the electron distribution function, $\varepsilon_{s_j}(\mathbf{k}_j)$ is the energy of an electron with wave vector $\mathbf{k}_j$. The two in the denominator of (7) appears because the final states are taken into account twice in the summation.

## 3. Results and discussion

### 3.1. Probability of impact ionization in CdHgTe quantum well

In this section, we present the results of impact ionization probability calculations for the B0225 structure grown in the group of N.N. Mikhailov from the Rzhanov Institute of Semiconductor Physics. Stimulated emission under optical excitation conditions was recently observed in this structure [25]. The structure contains 11 6.8 nm- $Cd_{0.067}Hg_{0.933}Te$ quantum wells separated by 30 nm $Cd_{0.7}Hg_{0.3}Te$ barriers. The structure was grown on the (013) plane. In the calculations, the temperature was assumed to be 10 K, what corresponded to the conditions for observing stimulated emission.

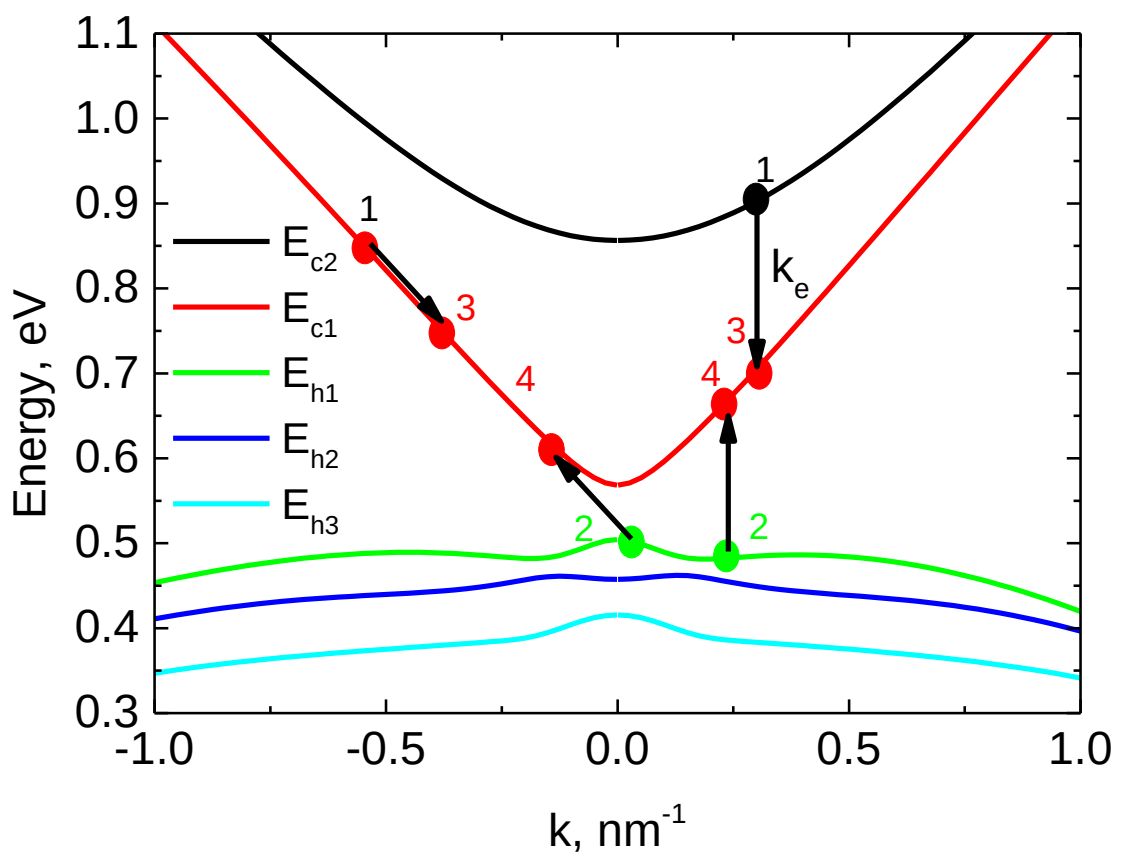

Fig.1. Band structure of a 6.8 nm $Cd_{0.067}Hg_{0.933}Te$ quantum well surrounded by $Cd_{0.7}Hg_{0.3}Te$ barriers. Arrows indicate possible electron transitions during impact ionization.

Figure 1 shows the band structure of the quantum well under consideration, calculated using the Kane model taking into account deformation defects. Details of the calculation can be found in reference [26]. In the calculation, the spin splitting of the subbands due to the absence of a lattice inversion center and the reduction of symmetry at the heteroboundaries were neglected, since these phenomena have little effect on the impact ionization in the structures under consideration. In total, the quantum well has three quantum-confinement subbands (ignoring the spin) in the conduction band and 14 quantum-confinement subbands in the valence band. The figure shows two subbands of the conduction band and three quantum-confinement subbands in the valence band, since the work considers impact ionization with the participation of electrons from these subbands. Figure 1 also shows the diagrams of electron transitions during impact ionization with a small change in the electron momentum (vertical arrows) and with a large change in the momentum (non-vertical arrows). From Figure 1 it is clear that vertical transitions are possible during the transition from the initial state of an electron located in the second subband of the dimensional quantum field to the final state in the first subband.

Let's discuss vertical transitions in more detail. The wave vector of the initial state of an electron in the valence band $\mathbf{k}_h$ for a given electron transition vector in the conduction band $\mathbf{k}_c$ is determined from the equation:

$$\varepsilon_{c2}(\mathbf{k}_c) - \varepsilon_{c1}(\mathbf{k}_c) = \varepsilon_{c1}(\mathbf{k}_{hj}) - \varepsilon_{hj}(\mathbf{k}_{hj}) \qquad (8)$$

where j=1,2,3 corresponds to the valence subband number. Figure 2 shows that as the value $\mathbf{k}_c$ increases, $\mathbf{k}_{hj}$ decreases, which is due to the convergence of the quantum confinement subbands in the conduction band.

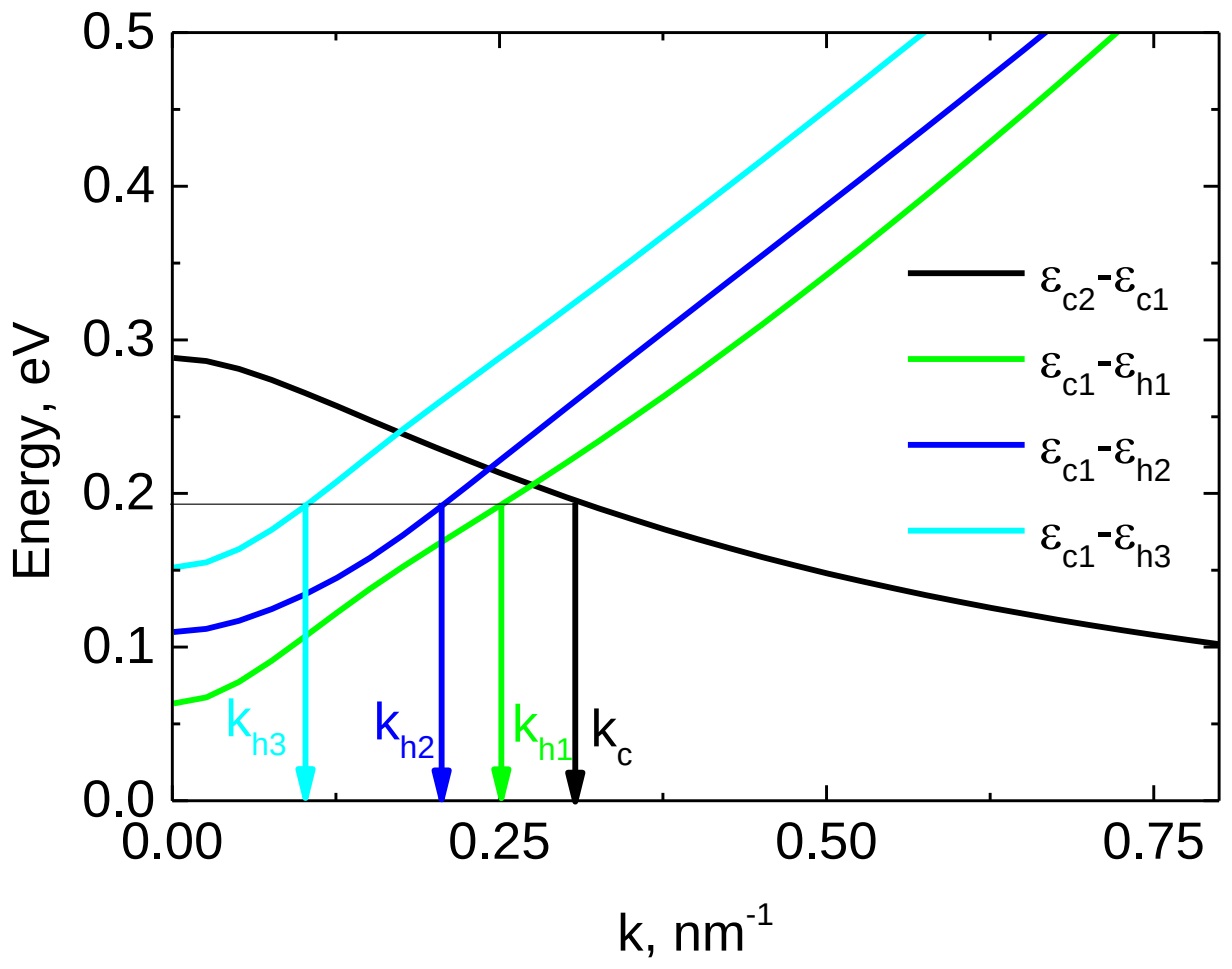

Fig. 2. Dependences of the difference in subband energies on the wave vector.

Obviously that a single value $\mathbf{k}_c$ corresponds to many values for $\mathbf{k}_{hj}$ a fixed j. All these values $\mathbf{k}_{hj}$ are located on a closed line in $\mathbf{k}$ space. For example, as Figure 3 shows, this line is close to a circle for $\mathbf{k}_{h1}$.

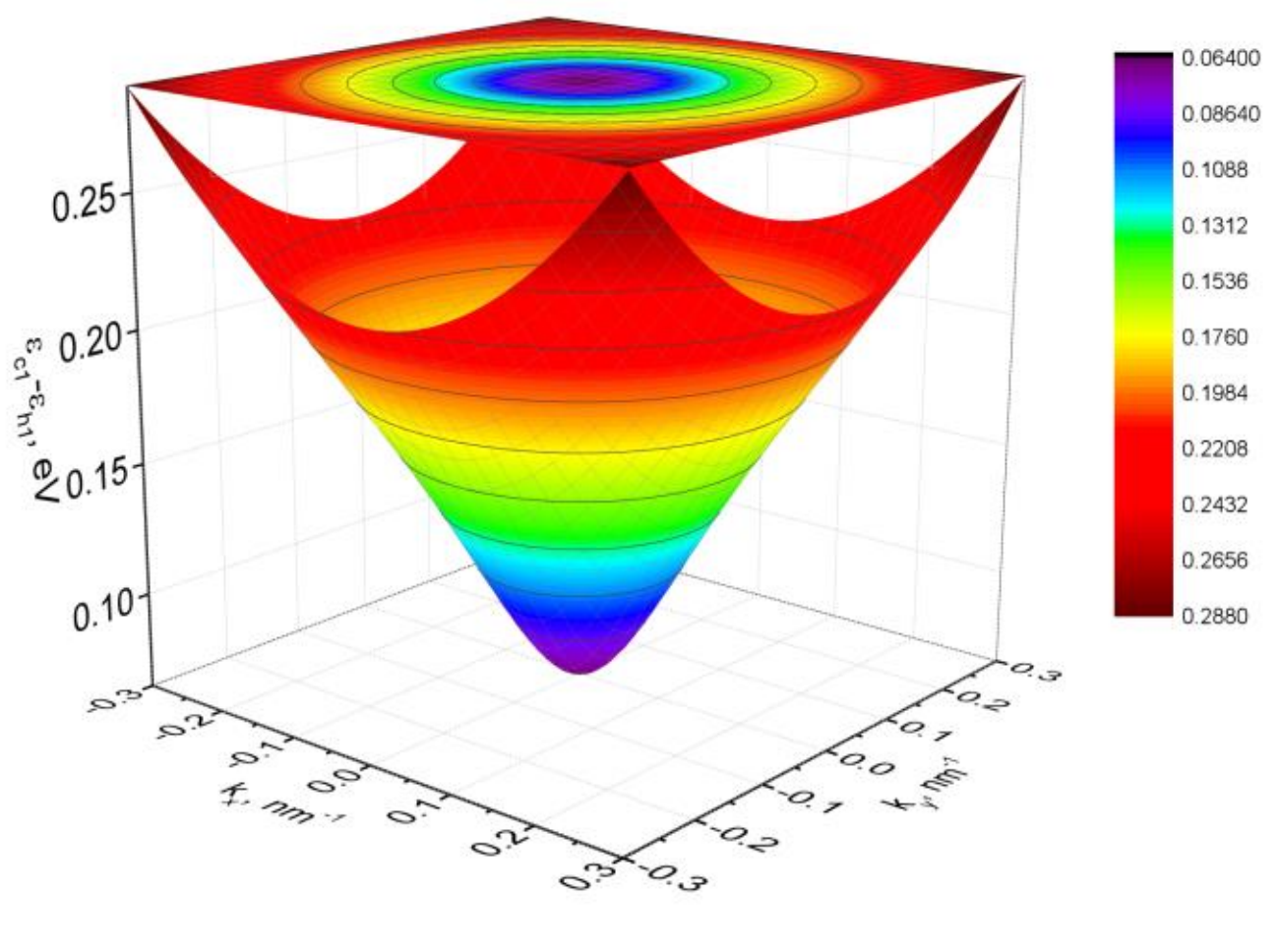

Fig. 3. Dependence $\varepsilon_{c1}(\mathbf{k}_{h1}) - \varepsilon_{h1}(\mathbf{k}_{h1})$ on $\mathbf{k}_{h1}$.

Figure 4 shows the calculated dependences of the impact ionization probabilities on the electron kinetic energy when the initial electron state in the conduction band is located in the first electron subband. The first two indices of the probability correspond to the subbands in which the initial electron states are located, and the remaining indices correspond to the final electron states. For example, $W_{c1h1c1c1}$ denotes the impact ionization probability when the initial electron states are in the first electron and hole subbands, and the final states are in the first electron subband. The quantity $\varepsilon_0$ is equal to the electron energy at the bottom of the first quantum confinement subband in the conduction band.

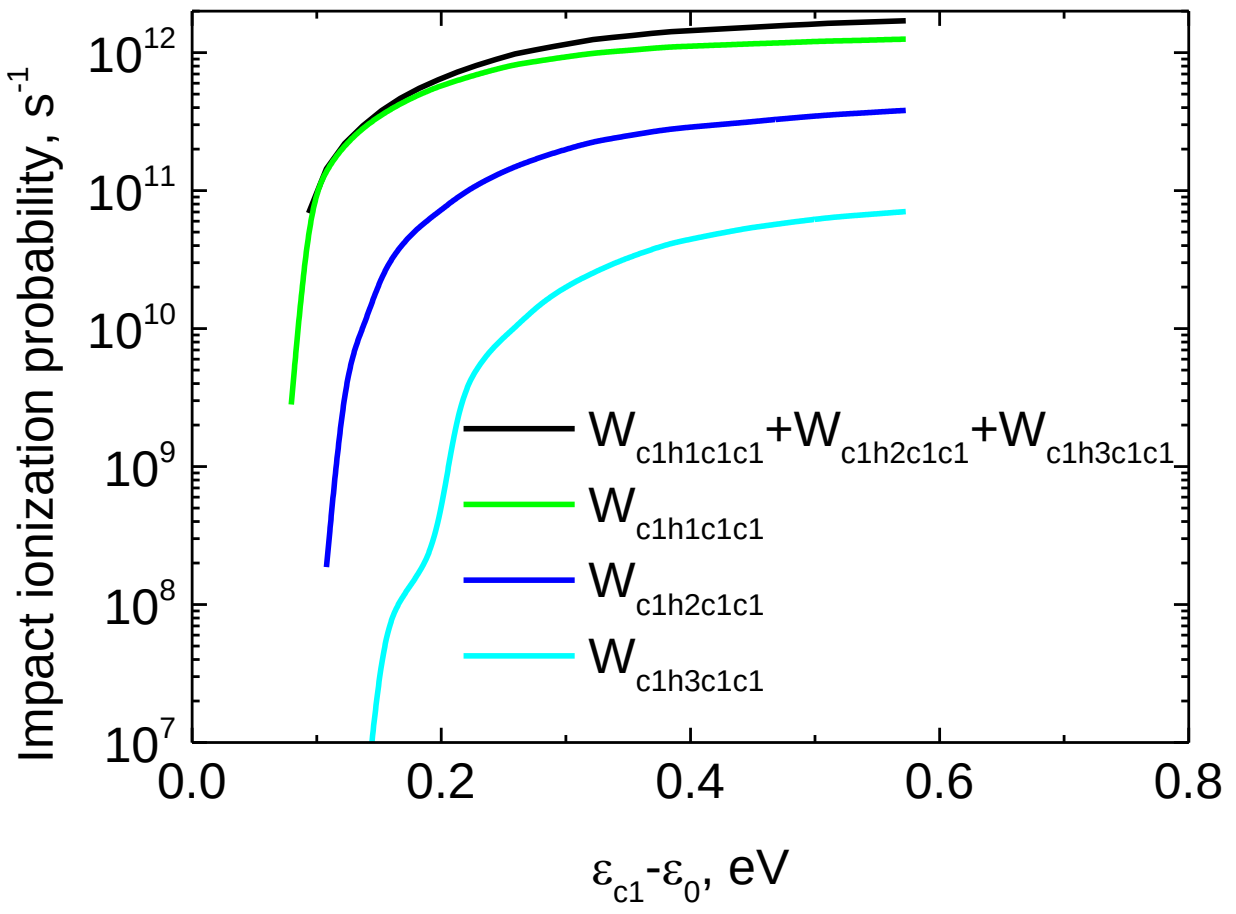

Fig. 4. Dependences of the impact ionization probabilities on the electron kinetic energy when the initial electron state in the conduction band is located in the first electron subband.

Figure 4 shows that as the number of the hole subband in which the electron's initial state resides increases, the probability of impact ionization decreases, while the threshold energy (the energy above which impact ionization is possible) increases. When calculating the probability of impact ionization, only three quantum-confinement subbands were considered, since the remaining subbands make a small contribution. The figure also shows that the total probability of impact ionization is close to the probability of impact ionization with electron ejection from the upper valence subband. Note that vertical impact ionization processes are impossible when the electron's initial state is in the lower subband of the conduction band. They are possible in the system under consideration if the electron's initial state is in the second subband of the conduction band.

Figure 5 shows the probabilities of impact ionization in the case where the initial state of an electron in the conduction band is in the second subband.

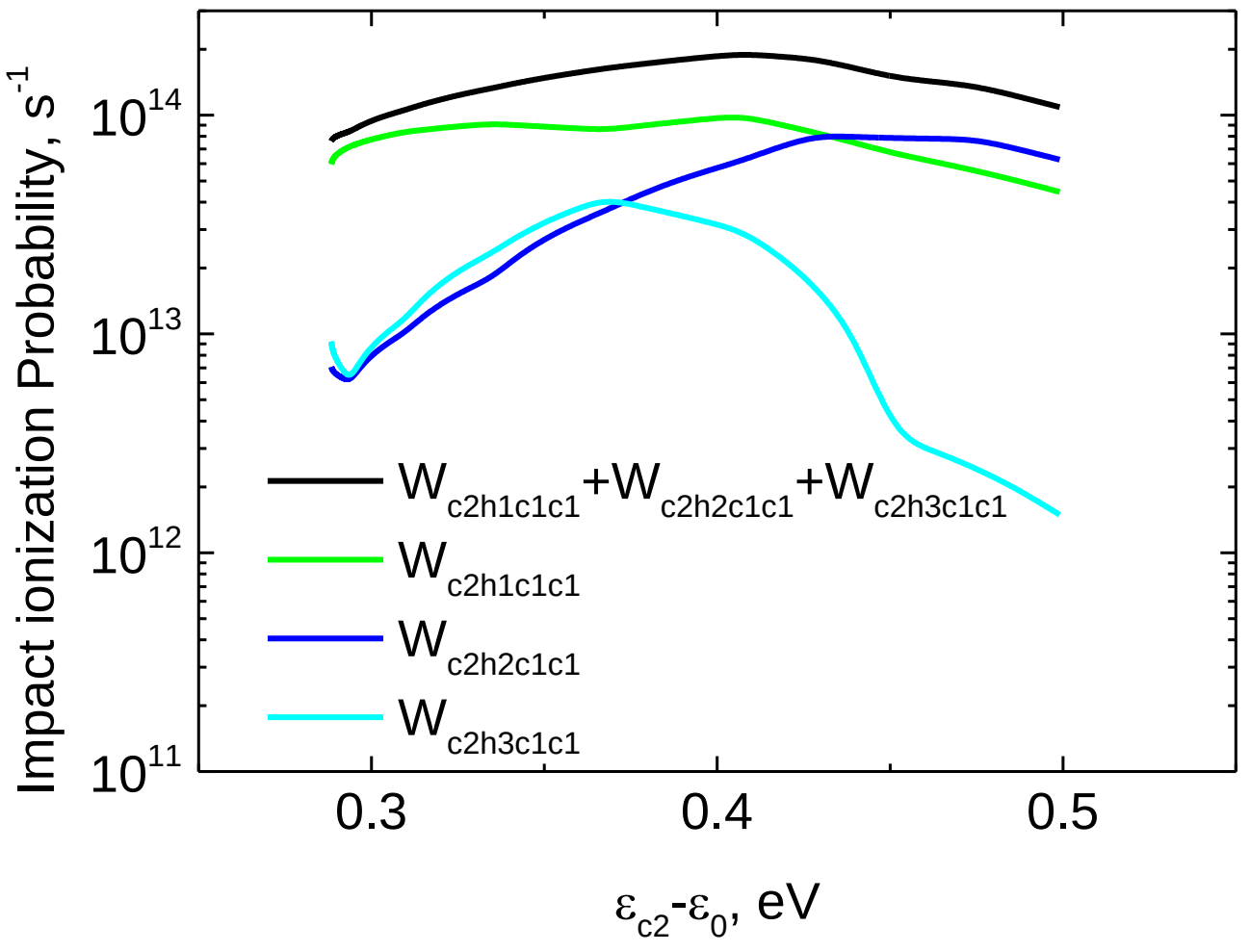

Fig. 5. Dependences of the impact ionization probabilities on the electron energy when the initial electron state in the conduction band is located in the second electron subband.

From Figure 5 it is evident that the probability of electron ejection from the third hole subband is significantly lower than from the first and second subbands. From a comparison of Figures 4 and 5 it is evident that the impact ionization probabilities for electrons in the first and second size-quantized subbands of the conduction band differ by almost two orders of magnitude. The reason for this is the impact ionization processes with a small change in the electron wave vector, which are possible for electrons in the second electron subband. This result is the main one in this work. It should be noted that a similar inequality holds in bulk GaSb semiconductors, where impact ionization processes with a small momentum transfer are possible (in this case, one of the final states of the electron is in the spin-split band of the electron state) [12].

The figure shows a non-monotonic dependence of the impact ionization probability on the electron energy, which is the result of competition between vertical and non-vertical transitions. The reason for the increase in the impact ionization probability with increasing energy is the contribution of non-vertical transitions, which increases monotonically with increasing electron energy (see Fig. 4). The reason for the decrease in the impact ionization probability with increasing electron energy is the decrease in the number of states participating in vertical transitions with increasing initial electron energy (see Figs. 1-3).

It should be noted that impact ionization, as a result of which the electron remains in the second subband of size quantization, has a much lower probability than impact ionization, the probabilities of which are presented in Figure 5. Therefore, such processes are not considered in this paper.

An energetic electron trapped in a quantum well as a result of optical excitation or capture from a barrier can lose energy not only through impact ionization but also through scattering by phonons. The most likely scattering process under the conditions under consideration is the emission of optical phonons.

This raises the question of the relationship between the probability of impact ionization and the probability of spontaneous emission of optical phonons. The next section presents the results of calculations of these probabilities.

### 3.2. Probability of optical phonon scattering

To find the spectrum of optical phonons and the potentials they create, we use the dielectric continuum model [27]. In this approximation, optical phonons are divided into two groups: bulk-like and surface. The frequency of bulk-like phonons coincides with the frequency of longitudinal optical phonons of the quantum well material. The frequencies of surface phonons are in the frequency ranges where the permittivities of the barrier and quantum well have opposite signs. Since the fraction of cadmium in the quantum well is small, for simplicity we used the permittivity of the HgTe lattice as the permittivity of the quantum well [28]:

$$\kappa(\omega) = \kappa_\infty \frac{\omega^2 - \omega_L^2}{\omega^2 - \omega_T^2} \qquad (9)$$

where $\kappa_\infty$ is the high-frequency permittivity of HgTe, $\omega_L$ and $\omega_T$ are the frequencies of the longitudinal and the transverse optical phonons in HgTe. For the permittivity of the barrier, the expression for the permittivity of the solid solution $Cd_xHg_{1-x}Te$ was used [29]:

$$\kappa_b(\omega) = \kappa_{\infty b} \frac{\left(\omega^2 - \omega_{LHgTe}^2\right)\left(\omega^2 - \omega_{LCdTe}^2\right)}{\left(\omega^2 - \omega_{THgTe}^2\right)\left(\omega^2 - \omega_{TCdTe}^2\right)} \qquad (10)$$

where $\kappa_{\infty b}$ is the high-frequency permittivity of $Cd_xHg_{1-x}Te$, $\omega_{LHgTe}$ and $\omega_{THgTe}$ are the frequencies of the longitudinal and transverse HgTe-like optical phonons in $Cd_xHg_{1-x}Te$, $\omega_{LCdTe}$ and $\omega_{TCdTe}$ are the frequencies of the longitudinal and transverse CdTe-like optical phonons in this material. The dependence of the longitudinal and transverse phonon frequencies on the composition of $Cd_xHg_{1-x}Te$ is given in [30].

Figure 6 shows the dependences of the energies of optical phonons on the wave vector.

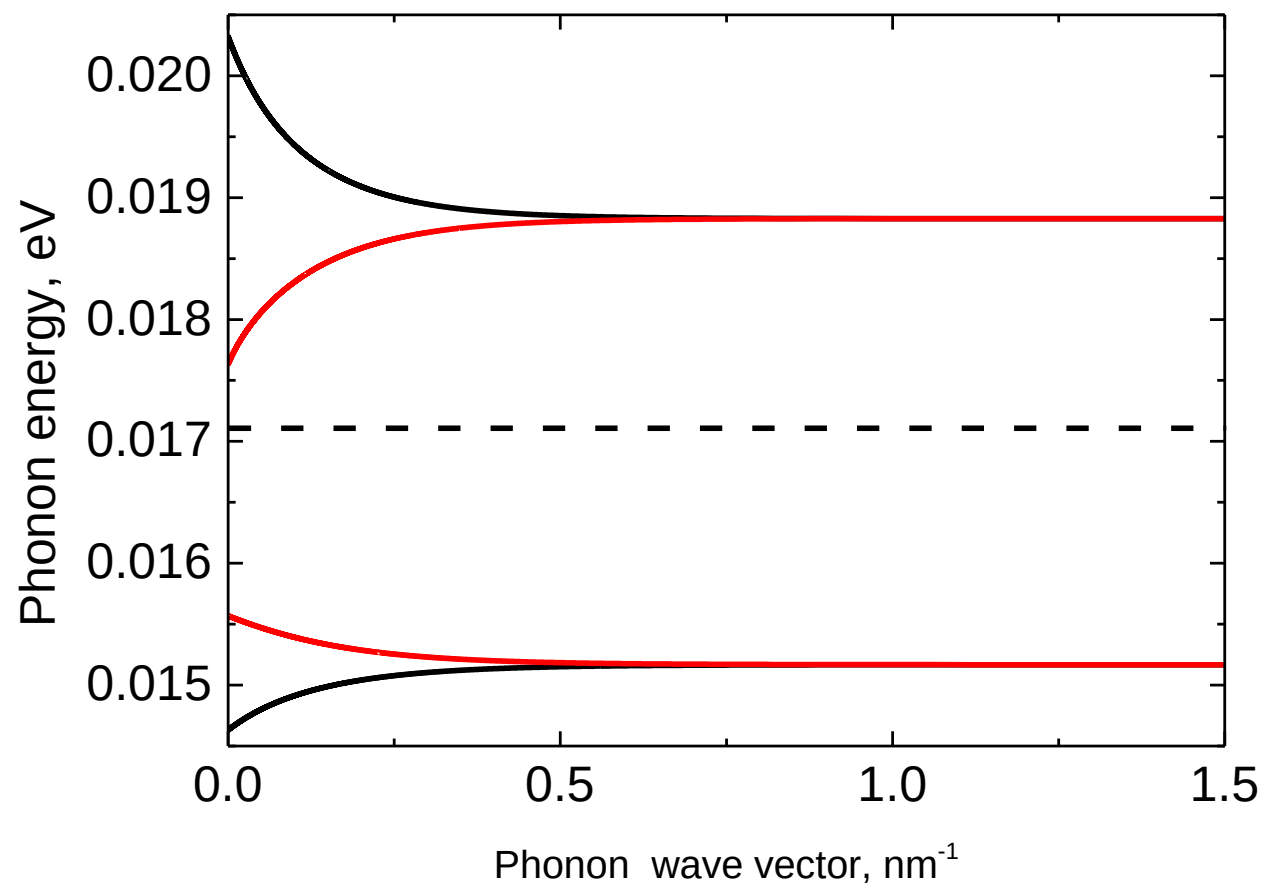

Fig. 6. Optical phonon spectra in a $Cd_{0.067}Hg_{0.933}Te/Cd_{0.7}Hg_{0.3}Te$ quantum well. The solid black lines correspond to even surface phonons, the potential generated by which is an even function relative to the quantum well center. The solid red lines correspond to odd optical phonons, the potential generated by which is an odd function relative to the quantum well center. The dashed line corresponds to bulk-like phonons.

The figure shows that there are four types of surface optical phonons in a quantum well: two symmetric and two antisymmetric.

Figure 7 shows the dependence of the probability of spontaneous emission of the optical phonons $W_{ij}^{ph}$ by an electron on the electron energy. The indices i and j denote the numbers of the subbands in which the electron was located before and after the emission of the optical phonon.

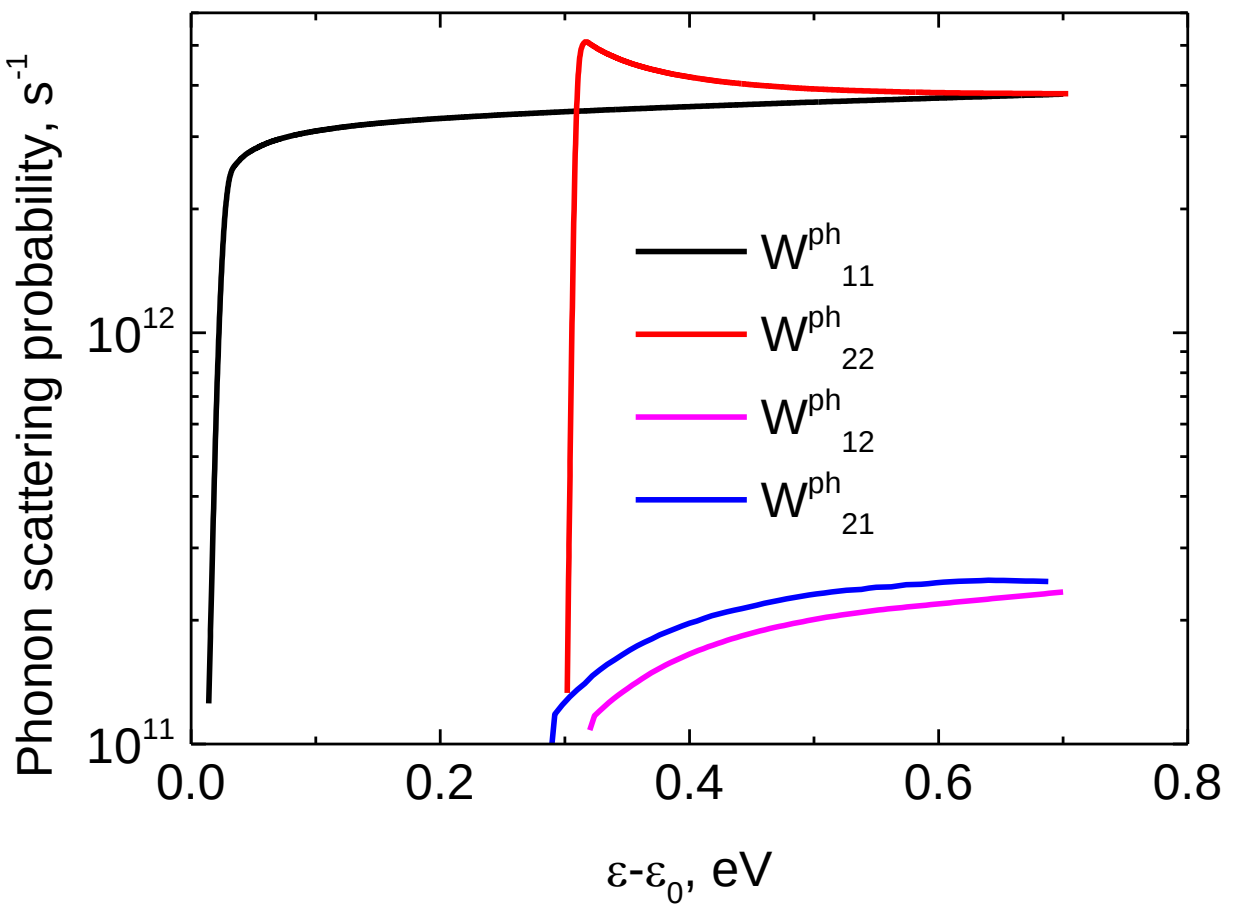

Fig. 7. Dependences of the optical phonon emission probabilities on the electron energy.

The figure shows that the probabilities of intersubband electron scattering by phonons are much lower than the probabilities of intraband scattering. Note that the largest contribution to intraband scattering comes from scattering by high-frequency even surface phonons.

Comparing Figures 4 and 5 with Figure 7, we can conclude that the probability of impact ionization caused by an electron in the second quantum confinement subband is more than an order of magnitude greater than the probability of optical phonon emission. However, the probability of optical phonon emission by an electron in the first quantum confinement subband is higher than the probability of impact ionization. Therefore, it is clear that an electron in the second subband will necessarily undergo impact ionization before losing energy to optical phonons.

A natural question arises: what is the probability of impact ionization caused by an electron from the first quantum confinement subband during its energy loss through the emission of optical phonons? To answer this question, we will estimate the probability of occurring once impact ionization caused by an electron from the first quantum confinement subband. To do this, we will make two simplifying assumptions for electrons with energies greater than the threshold energy for impact ionization. First, we will assume that the electron loses energy continuously, not in portions equal to the optical phonon energy. We will assume that the rate of energy loss by an electron due to phonon emission is $\hbar\omega_0 W_{11}(\varepsilon)$, where $\hbar\omega_0$ is the optical phonon energy and $W_{11}^{ph}(\varepsilon)$ is the probability of optical phonon emission. We will take the energy of an even high-energy surface phonon as $\hbar\omega_0$, since scattering on it is most probable. Secondly, in the equations for the electron distribution function, we will neglect the term for the arrival due to impact ionization (since we are considering only the process of occurring once impact ionization). We will consider a steady-state situation where electrons enter the first subband of size quantization due to photoexcitation with energy $\varepsilon_i$ and rate $j_0$. In this case, the following equation is valid for the electron distribution function $f(\varepsilon)$:

$$\frac{dj(\varepsilon)}{d\varepsilon} = -W(\varepsilon)g(\varepsilon)f(\varepsilon) \tag{11}$$

where $j(\varepsilon) = -\hbar\omega_0 W_{11}(\varepsilon)g(\varepsilon)f(\varepsilon)$ is the electron flux in the energy space, $W(\varepsilon) = W_{c1h1c1c1}(\varepsilon) + W_{c1h2c1c1}(\varepsilon) + W_{c1h3c1c1}(\varepsilon)$ is the probability of impact ionization of an electron in the first subband with energy $\varepsilon$, and $g(\varepsilon)$ is the electron density of states. Note that the quantity $j(\varepsilon)$ is negative because the electron is cooling, i.e., losing its energy due to interaction with optical phonons; therefore, the quantity $j_0$ is also negative. The probability of occurring once impact ionization when an electron cools from the energy $\varepsilon_i$ to the threshold that is:

$$w(\varepsilon_i) = \frac{1}{j_0}\int_{\varepsilon_{th}}^{\varepsilon_i} d\varepsilon' W(\varepsilon')g(\varepsilon')f(\varepsilon') \tag{12}$$

where $\varepsilon_{th}$ is the impact ionization threshold energy for an electron in the first subband. Integrating Eq.(11), we obtain $w(\varepsilon_i) = (j(\varepsilon_i) - j(\varepsilon_{th}))/j_0$. Equation (11) can be rewritten as:

$$\frac{dj(\varepsilon)}{d\varepsilon} = \frac{W(\varepsilon)}{\hbar\omega_0 W_{11}(\varepsilon)} j(\varepsilon) \tag{13}$$

Integrating Eq. (13) we find:

$$j(\varepsilon) = j_0 \exp\left( \int_{\varepsilon_i}^{\varepsilon} d\varepsilon' \frac{W(\varepsilon)}{\hbar\omega_0 W_{11}(\varepsilon)} \right), \qquad \varepsilon < \varepsilon_i \tag{14}$$

Using Eq. (14), we find the probability of occurring once impact ionization:

$$w(\varepsilon_i) = 1 - \exp\left( -\int_{\varepsilon_{th}}^{\varepsilon_i} d\varepsilon' \frac{W(\varepsilon)}{\hbar\omega_0 W_{11}(\varepsilon)} \right) \tag{15}$$

Figure 8 shows the dependence $w(\varepsilon_i)$.

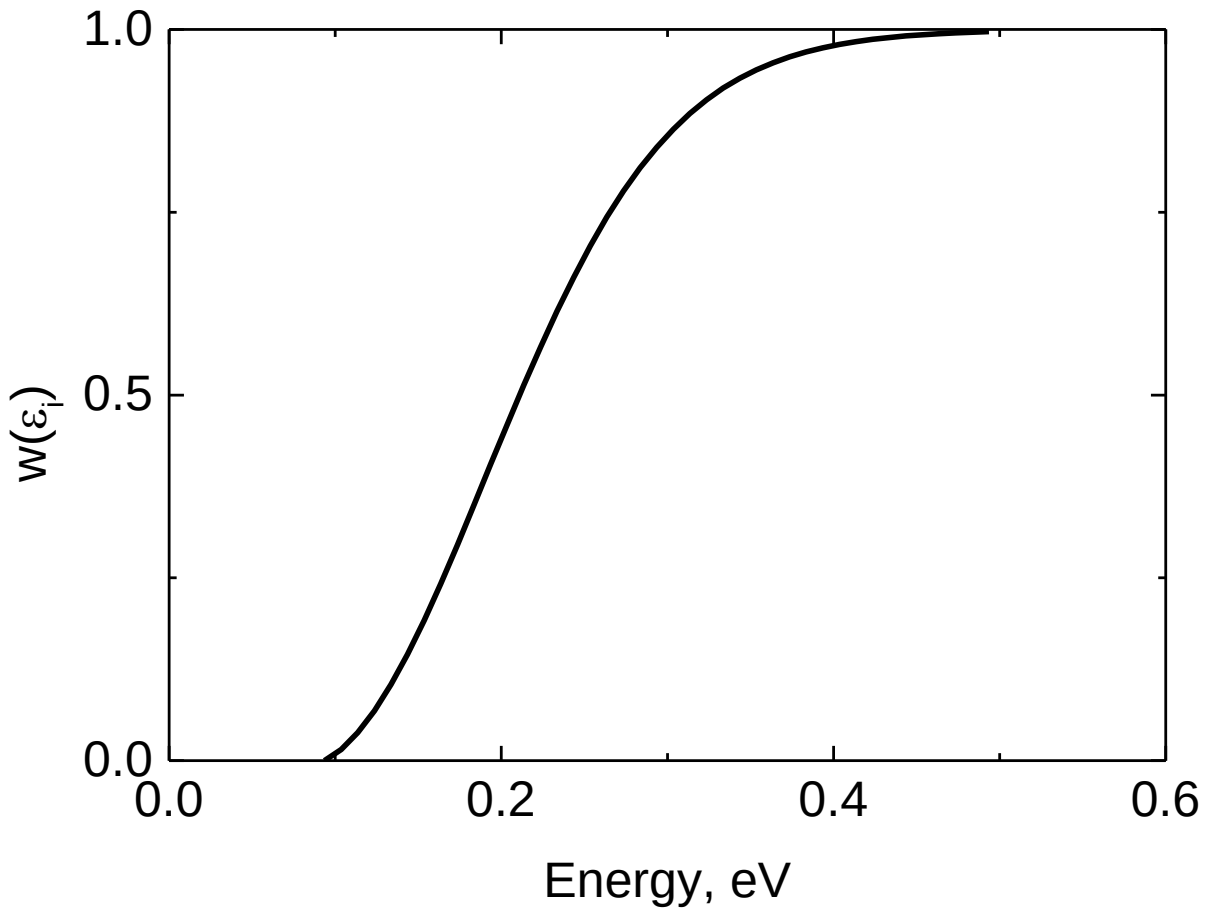

Fig. 8. Dependence of occurring once impact ionization on the initial energy of an electron "cooling" in the first electron subband

It should be noted that impact ionization can be caused by an electron that has already participated in impact ionization one or more times. A discussion of the probability of such processes is beyond the scope of this paper.

**Conclusion**

In conclusion, we present the main results of the study and discuss the methods for observing impact ionization in the structure under consideration. In this study, we calculated the dependences of the impact ionization probability on the electron energy in the conduction band of a 6.8 nm $Cd_{0.067}Hg_{0.933}Te$ quantum well surrounded by $Cd_{0.7}Hg_{0.3}Te$ barriers. It was shown that the probability of impact ionization caused by an electron from the second subband is approximately two orders of magnitude higher than a similar process for an electron from the first subband. The reason for this is in electron transitions with a small change in the wave vector, which are possible in this structure due to the fact that the distance between the first and second subbands in the conduction band is close to the band gap. A comparison of the impact ionization probabilities and the probabilities of spontaneous emission of optical phonons was performed. The dependence of the probability of occurring once impact ionization on the initial energy of an electron in the first subband during its energy loss due to the emission of optical phonons was found.

Let us now discuss methods for observing impact ionization in the structure under study under optical excitation. To this end, we note that the transmission of light through the structure with phonon energies smaller than the barrier band gap is determined by the absorption coefficient of the quantum wells, while the photoconductivity depends on the photocarrier concentration, which is contributed by impact ionization. By comparing the transmission and photoconductivity spectra, we can detect impact ionization processes. Note that the dependences of the integrated photoluminescence intensity or the threshold laser generation intensity on the photon energy of the excitation radiation can be used instead of photoconductivity.

**ACKNOWLEDGMENTS**
The impact ionization calculation was supported by the Russian Science Foundation (Project No. 22-12-00310-П), the phonon scattering calculation was carried out within the state assignment of Ministry of Science and Higher Education of the Russian Federation (theme No. 124050300055-9/FFUF-2024-0045).

## AUTHOR DECLARATIONS

Conflict of Interest

The authors have no conflicts to disclose.

### Author Contributions

V. Ya. Aleshkin: Conceptualization (lead); Data curation (equal); Formal analysis (equal); Funding acquisition (equal); Investigation (equal); Methodology (equal); Software (equal); Validation (equal); Visualization (equal); Writing – original draft (equal); Writing –review & editing (equal).
A. A. Dubinov: Data curation (equal); Formal analysis (equal); Funding acquisition (equal); Investigation (equal); Methodology (equal); Validation (equal); Visualization (equal); Writing – original draft (equal); Writing – review & editing (equal).
V. V. Rumyantsev: Conceptualization (equal); Data curation (equal); Formal analysis (equal); Funding acquisition (equal); Investigation (equal); Methodology (equal); Visualization (equal); Writing – original draft (equal); Writing – review & editing (equal).

## DATA AVAILABILITY

The data that support the findings of this study are available
from the corresponding author upon reasonable request.

**Appendix**: **Coulomb potential of an electron in a quantum well.**

For a quantum well occupying the region $|z|<d/2$, the potential can be represented as follows for $|z_1|<d/2$:

$$\varphi_q(z_1,z_2)=\begin{cases} \dfrac{2\pi e}{q\kappa_2}\left[\exp(-q|z_2-z_1|)+\dfrac{\exp(qz)\delta\left[\exp(q(d-z_1))+\delta\exp(-q(z_1+d))+(\delta+1)\exp(qz_1)\right]}{\exp(qd)-\delta^2\exp(-qd)}\right], & z_2<-d/2 \\ \dfrac{2\pi e}{q\kappa_2}\left[\exp(-q|z_2-z_1|)+\dfrac{2\delta\left[ch(q(z_2+z_1))+\delta\exp(-qd)ch(q(z_2-z_1))\right]}{\exp(qd)-\delta^2\exp(-qd)}\right], & |z_2|<d/2 \\ \dfrac{2\pi e}{q\kappa_2}\left[\exp(-q|z_2-z_1|)+\dfrac{\delta\left(\exp(q(z_1+d))+\delta\exp(q(z_1-d))+(\delta+1)\exp(-qz_1)\right)}{\exp(qd)-\delta^2\exp(-qd)}\right], & z_2>d/2 \end{cases}$$

(A1)

where $\delta=\dfrac{\kappa_2-\kappa_1}{\kappa_2+\kappa_1}$, and $\kappa_{1,2}$ are high-frequency permittivities of the quantum well and barrier, respectively.

For $z_1<-d/2$:

$$\varphi_q(z_1,z_2)=\begin{cases} \dfrac{2\pi e}{q\kappa_1}\exp(q|z_1-z_2|)-\dfrac{2\pi e}{q\kappa_1}\dfrac{\delta\exp(qz_1+qd+qz_2)sh(qd)}{\exp(qd)-\delta^2\exp(-qd)}, & z_2<-d/2 \\ \dfrac{2\pi e}{q\kappa_1}\exp(q|z_1-z_2|)+\dfrac{\pi e\delta}{q\kappa_1}\dfrac{\exp(qz_1)\left[\delta\exp(q(-d-z_2))-\exp(q(d-z_2))+(1-\delta)\exp(qz_2)\right]}{\exp(qd)-\delta^2\exp(-qd)}, & |z_2|<d/2 \\ \dfrac{2\pi e}{q\kappa_1}\exp(q|z_1-z_2|)-\dfrac{2\pi e\delta^2}{q\kappa_1}\dfrac{\exp(q(z_1-z_2))sh(qd)}{\exp(qd)-\delta^2\exp(-qd)}, & z_2>d/2 \end{cases}$$

(A2)

For $z_1>d/2$

$$\varphi_q(z_1,z_2)=\begin{cases}\frac{2\pi e}{q\kappa_1}\exp(-q|z_2-z_1|)-\frac{2\pi e}{q\kappa_1}\frac{\delta^2\exp(q(z_2-z_1))sh(qd)}{\exp(qd)-\delta^2\exp(-qd)}, & z<-d/2\\ \frac{2\pi e}{q\kappa_1}\exp(-q|z_2-z_1|)+\frac{2\pi e\exp(-qz_1)}{q\kappa_1}\frac{\delta\left[\delta\exp(q(z_2-d))+(1-\delta)\exp(-qz)-\exp(qd+qz_2)\right]}{\exp(qd)-\delta^2\exp(-qd)}, & |z|<d/2\\ \frac{2\pi e}{q\kappa_1}\exp(-q|z_2-z_1|)-\frac{2\pi e\delta}{q\kappa_1}\frac{\exp(qd-qz_1-qz)sh(qd)}{\exp(qd)-\delta^2\exp(-qd)}, & z>d/2\end{cases} \quad \text{(A3)}$$

Using Eq. (6) and Eqs. (A1-A3) we obtain for $|z_1|<d/2$

$$g(z_1,z_2)=\begin{cases}\frac{2\pi e}{\kappa_1}\left[\frac{d}{2}\left(1-\frac{\kappa_2}{\kappa_1}\right)+z_2-\frac{\kappa_1}{\kappa_2}z_1\right], & z_2<-d/2\\ \frac{2\pi e}{\kappa_2}\left[\frac{d}{2}\left(1-\left(\frac{\kappa_2}{\kappa_1}\right)^2\right)-|z_2+z_1|\right], & |z_2|<d/2\\ \frac{2\pi e}{\kappa_1}\left[\frac{d}{2}\left(1-\frac{\kappa_2}{\kappa_1}\right)-z_2+\frac{\kappa_1}{\kappa_2}z_1\right], & z_2>d/2\end{cases} \quad \text{(A4)}$$

For $z_1<-d/2$

$$g(z_1,z_2)=\begin{cases}\frac{2\pi e}{\kappa_1}\left(\frac{d}{2}\left(\frac{\kappa_1}{\kappa_2}-\frac{\kappa_2}{\kappa_1}\right)-|z_2-z_1|\right), & z_2<-d/2\\ \frac{2\pi e}{\kappa_1}\left(\frac{d}{2}\left(1-\frac{\kappa_2}{\kappa_1}\right)+z_1-z_2\frac{\kappa_1}{\kappa_2}\right), & |z_2|<d/2\\ \frac{2\pi e}{\kappa_1}\left(z_1-z_2-\frac{d}{2}\frac{\kappa_1}{\kappa_2}\left(1-\frac{\kappa_2}{\kappa_1}\right)^2\right), & z_2>d/2\end{cases} \quad \text{(A5)}$$

For $z_1>d/2$

$$g(z_1,z_2)=\begin{cases}\frac{2\pi e}{\kappa_1}\left[d\left(1-\frac{\kappa_1}{2\kappa_2}-\frac{\kappa_2}{2\kappa_1}\right)+z_2-z_1\right], & z_2<-d/2\\ \frac{2\pi e}{\kappa_1}\left[\frac{d}{2}\left(1-\frac{\kappa_2}{\kappa_1}\right)+\frac{\kappa_1}{\kappa_2}z_2-z_1\right], & |z_2|<d/2\\ \frac{2\pi e}{\kappa_1}\left[\frac{d}{2}\left(\frac{\kappa_1}{\kappa_2}-\frac{\kappa_2}{\kappa_1}\right)-|z_2-z_1|\right], & z_2>d/2\end{cases} \quad \text{(A6)}$$